# Potentials and Limitations on Different Busbar Protections in Industrial Applications

**Karl M.H. LAI†, Yunhe HOU* and Kwunhang WONG****

**Abstract** – Busbar protection is a cornerstone of industrial power system reliability, as failures at switchgear can propagate rapidly and extend restoration times. The Taiwan "303 blackout" in 2022, initiated by a short-circuit fault and aggravated by CT supervision lockout, illustrates the severe consequences of inadequate designs on busbar protection in centralized grids.

Despite extensive academic and industrial guidance, practical challenges remain in busbar protection design, including CT placement, dynamic zone selection, CT saturation, and evolving fault scenarios. Standards and guides often fail to reflect operational limitations under diverse system conditions.

This paper provides a structured evaluation of busbar protection (BBP) schemes in industrial applications. Leakage-to-frame, high-impedance differential (87Z), interlocking overcurrent (ILOC), electronic (EBBP), and numerical (NBBP) schemes are examined with respect to their operating principles, implementation requirements, and suitability for different bus configurations. Comparative analysis highlights clear trade-offs in different busbar protection schemes.

The main contributions are: (i) systematic review of classical and modern busbar protection schemes, (ii) identification of operational limitations with various busbar protection, and (iii) comparison on the integration of CT supervision, check zones, and directional checks in NBBP to mitigate maloperations. The discussion reinforces the role of robust busbar protection design in preventing cascading outages and maintaining system reliability.

## 1. Introduction – Busbar Protection as a Critical Substation Protection

Busbar protection (87B) is an essential element of switchgear protection, with its protection zone overlapping the zones of all connected circuits. The busbar serves as a node in power systems connecting lines, transformers and all other ancillary equipment including reactive power compensation devices such as capacitor banks, reactors and condensers, and possibly earthing equipment such as earthing transformers. A single failure on the busbar possibly leads to a prolonged outage which may heavily affect system reliability. The outage duration depends on busbar arrangements, gas zones, and current transformer (CT) locations. Besides, busbar protection is often equipped with breaker failure protection (BFP), which is a local backup to the circuit protection to clear the fault by clearing the whole busbar or tripping the source under breaker stuck conditions or trip circuit failure

This highlights the importance of well-designed busbar protection in maintaining system reliability and enabling selectivity and dependability within protection schemes.

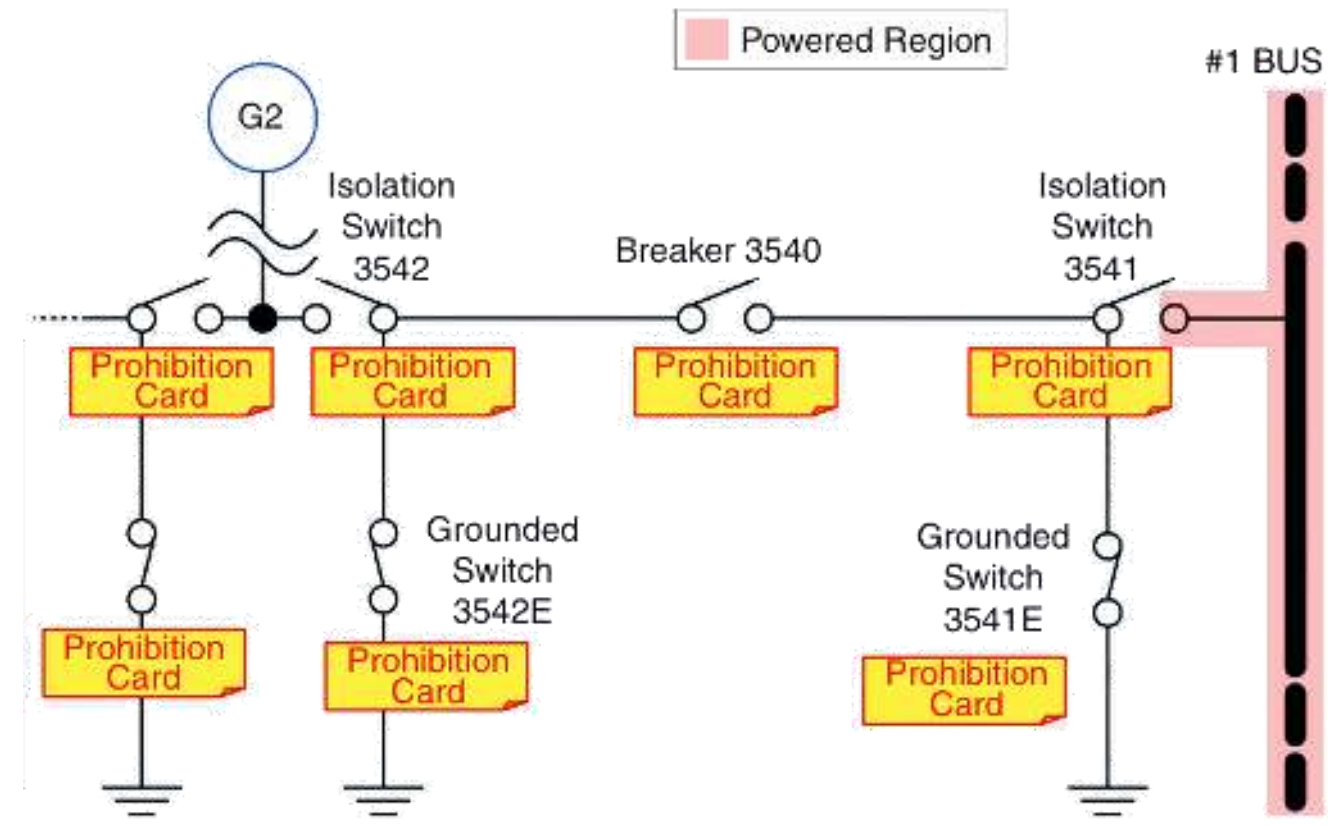

**Fig. 1.** Taiwan 303 Blackout Incident in 2022 [1]

† Corresponding Author : Dept. of ECE, The University of Hong Kong & CLP Power Hong Kong (perspect@connect.hku.hk)

* Dept. of Electrical and Computer Engineering, University of Hong Kong (yhhou@eee.hku.hk)

** Dept. of Electrical and Computer Engineering, University of Hong Kong (u3556440@connect.hku.hk)

Taiwan 303 blackout in 2022 was caused by operational errors and design limitations of busbar protection [1]. Under a generator overhaul in Xingda power plant with the isolation-and-earth location as illustrated in **Figure 1**, the $SF_6$ gas in the gas zone of circuit breaker 3540 has been removed and the isolation switch 3541 was locked for an interlock replacement. After the work for isolator 3541, the operational person requested to perform a function check on isolator 3541. The section between isolator 3541 and circuit breaker 3540, which was with insufficient insulation, was then energized onto fault.

With seven seconds closing for isolator 3541, the busbar differential detected a small but evolving fault current, which was considered as an open circuit condition, hence the busbar protection automatic switched out (lockout) itself five seconds after the fault inception to maintain its stability. Since the switchyard was interconnected without sectionalization, this generator-outlet fault was then cleared by the remote back-up protection, i.e. distance Zone 2, and the disconnection of all outgoing feeders. The loss of whole generation substation led to a grid partition and blackout upon generator over-frequency (OF) trip at generator side and under frequency load shedding (UFLS) at load side.

The fundamental principles of busbar protection (87B) are based on Kirchoff's current law (KCL), to evaluate the phasor or instantaneous sum entering and leaving the node; directional overcurrent (DOC), which identifies the fault current flow into the zone; or simply OC which indicates a flashover to the enclosure through a CT to ground.

High Impedance Busbar Protection (HZBBP), Electronic Busbar Protection (EBBP), and Numerical Busbar Protection (NBBP) mainly rely on differential current calculations. If necessary, a restraint current can also be determined for a bias differential. HZBBP drives the differential current into the high impedance circuit (87Z) to activate a trip signal or a block signal. Electronic busbar protection, namely RADSS (Redundant Analog Digital Substation System) in ABB, is to drive the input current through a diode network to restraining path and differential path to provide restraint blocking and enable fast operation under internal fault. Modern busbar protection with NBBZ determines the differential current aided with check zone (CZ), external fault detectors (EFD), CT supervision function, cross stabilization to maintain stability under normal condition and enable fast tripping for internal fault without delays [2], [5].

Interlock Overcurrent (ILOC) employs directional (DIR) elements to determine fault current flow direction, i.e. from transformers to busbar, and trip if no other overcurrent (OC) element at downstream activated to indicate a downstream fault in the outgoing circuits [7], [20], [21]. It trips the source to clear the fault instead of clearing the whole bus. Leakage-to-Frame (LTF) employs OC to detect fault current flowing between the switchgear frame earth and the true earth.

Although the underlying principles are straightforward, applications are complex due to the following challenges.

**CT saturation** under external fault conditions necessitates intrinsic stability at the protection schemes - such as the small shunt voltage in HBBP with the reverse voltage measured, the large bias voltage in EBBP or the speed requirement for EFD blocking in NBBP [2], [4]. **Dynamic bus replica** is to determine the zone the current to be counted and the zone to trip with the auxiliary status, which may not truly reflect the actual position of the primary contact due to the stochastic flashover time and mechanism performance [4]. It could even lead to spurious differential and false trips, especially under on-load transfer conditions [2], [5]. **CT failure detection** is to avoid tripping the whole bus with loads of several MWs due to issues such as open CT or CT saturation. **Expandability** of busbar protection is to guarantee no large changes in secondary and hence prolonged outage under busbar extension. **Evolving faults**, an external fault evolving into an internal fault or a small fault current growing into a large fault, are often cleared with a delay to avoid spurious tripping, which is to be overcome with the aids of adaptive logic with DIR or EFD in NBBP [4].

Busbar protection in sum is to clear faults occurring within the switchgear measuring zone and to serve as a local backup for feeder protection if necessary. Despite its importance, the limitations of various schemes are not comprehensively documented in the literature. In particular, the ILOC scheme, which relies on directional elements and has limited global application, and the NBBP, whose operating principles vary significantly among different manufacturers, remain insufficiently characterized in terms of performance constraints and practical deployment challenges.

This paper presents a comprehensive review of the limitations associated with both classical and modern busbar protection schemes under diverse operating scenarios, supported by industrial experience worldwide. The constraints of the LTF scheme are first outlined to demonstrate the necessity of implementing ILOC protection with its inherent limitations in **Section 2**. The operating principles of HZBBP and the evolution of the BFP scheme are then discussed in **Section 3**. **Section 4** addresses the application of EBBP, highlighting its limitations under varying system configurations and its susceptibility to circuit or component failures. Finally, **Section 5** compares the designs of NBBP implemented by different manufacturers with classical schemes, providing a consolidated assessment of their effectiveness in addressing the challenges identified.

## 2. From Leakage-to-Frame (LTF) to Interlock Overcurrent (ILOC)

Leakage-to-Frame Busbar Protection, as coined Faulted Bus Scheme in IEEE std. C37.234 [6], has been widely used for high voltage (11kV ~ 33kV) busbar protection with its simple design. The metal-clad switchgear enclosure is insulated from the ground and connected through a fault-bus CT to the true earth [6]. The scheme can be designed stable with additional check CTs installed at transformer neutral and designed selective with additional directional check at bus coupler and bus section, which was known as zone-discriminative LTF (ZD-LTF). However, the performance of LTF was limited with –

- degraded frame-to-earth insulation, resulting in reduced protection sensitivity.
- limited detection to earth-fault conditions, with no response to other fault types.
- potential maloperation under cable end box (CEB) faults, particularly if a flashover occurs to the frame.
- possible maloperation during through-fault conditions arising from improper cable sheath connections, which should remain insulated from the frame.

Hence, LTF is often replaced with another busbar protection scheme, especially with a degraded frame-earth insulation. The possible option is often HZBBP, yet it requires dedicated Class X CTs with racking switches, which is to indicate the racked position for the circuit breaker (CB) in Air Insulated Switchgear (AIS) or the busbar isolator auxiliary contacts in its counterpart, and a trip relay in each panels. A possible option is Rough Balance Scheme [19], a.k.a. Partial Differential Scheme in [6], which is not preferrable with the delay and poor sensitivity in operation. Another simpler option is ILOC, which only requires a dedicated panel with secondary modifications on the CT wiring and on the additional blocking and tripping relays.

ILOC, as known as Directional Blocking Zone-Interlock Scheme in IEEE std. C37.234 [6], is intrinsically a zone-discriminative scheme to clear a fault by tripping its source, and providing a BFP function. In a typical distribution network as illustrated in **Figure 2(a)**, the busbar is connected to the source, possibly transformers in parallel or interconnectors, and the loads. Fault current flows from the source to the bus, or through the bus to the load ends. If the fault current flows into the bus, as a "reverse" direction as depicted in **Figure 2(a)**, and there is no blocking signal issued from the outgoing circuits, such as feeders or capacitor banks, or from the paralleled transformer with forward DIR as an indication of HV side fault, the fault is trusted as a busbar fault.

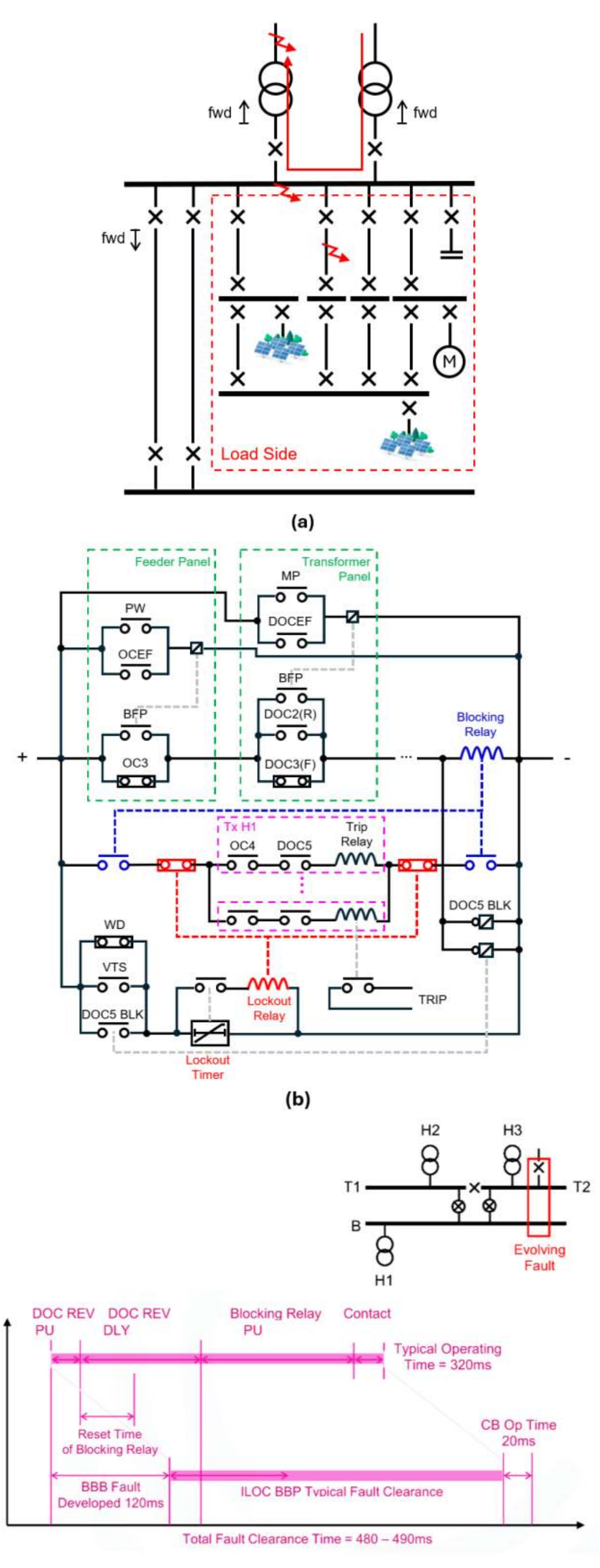

**Fig. 2.** (a) Typical 11kV Distribution Network
(b) Possible Hard-wire Implementation of ILOC Scheme
(c) Evolving Fault with ILOC

One possible hard-wire implementation which relies on DOC and OC contacts of BBP relays for ILOC scheme is as depicted in **Figure 2(b)**.

When there is a busbar fault, the source CT senses a reverse current such that the OC4 contact (as an OC guard) and the DOC5 contact (as a DIR check) closes after a time delay to clear the fault. The time delay is set to ensure the operation of OC3 contact at load side can successfully block the scheme operation in case of downstream fault. It is by de-energizing the blocking relay with an N.O. contact to disconnect the DC to the tripping element (pink box).

Under a feeder fault at a legged ring, the BI to enable DOC5 is temporarily de-energized such that the DOC5 contact cannot operate even with a reverse current. Similarly, when there is a forward fault, i.e. a fault occurring at the HV side of the paralleled transformer, DOC3 contact opens to de-energize the blocking relay. To provide a BFP function, one can energize the BFP binary input (BI) to activate the BFP contact to complete the blocking path after the internal breaker failure timer delay (BFT). It is noted that a phase-to-phase busbar fault can still open the DOC3 contact as the fault current flows from one phase into the bus and returns back as a forward current at the phase. To avoid any instability, the scheme is switched lockout to disconnect the DC if any of the busbar protection (BBP) relay is defective with the watchdog (WD) contact or the voltage transformer supervision (VTS) contact activates.

From the above discussion, one can realize the limitations of such scheme from the network in **Figure 2(a)**. First, continuous or transient overcurrent into the bus can be considered as a fault to be cleared in this directional overcurrent scheme. During emergency conditions when maximum cyclic rating (MCR) of a cable can be utilized as a final network reserve to feed the load, the DOC considers it as an in-zone fault, with no blocking signal to be activated at downstream, and the scheme clears this overcurrent by tripping the source. Stalling current after fault or any voltage disturbance is also a reverse OC. Therefore, a temporary setting change is also required under such condition. Second, this scheme relies on a clear definition on source and load. With the growth of distributed energy resources (DERs), the DERs can lead to a bidirectional or circulating flow in the 11kV ring network. It should be considered as a source to be cleared under a busbar fault. Yet, it is difficult to put all circuits into a source configuration as additional voltage selection scheme is to be installed to provide a polarizing signal to the directional elements and it could lead to continuous blocking under temporary overload or stalling current conditions in which a forward overcurrent can be seen. The DERs under loss-of-main condition can also be cleared with anti-islanding logic, possibly with an active signal by transfer trip or a passive logic with voltage vector shift (VVS) or rate-of-change of frequency (ROCOF) by the DERs. The effect of DERs to ILOC was further discussed in [8]. Third, an evolving earth-fault with ILOC can lead to a delayed fault clearance. As illustrated in **Figure 2(c)**, a fault at top busbar (TBB) in an AIS can evolve to a double busbar fault as the two buses are in same enclosure. The blocking signal to maintain stability of the scheme asserted in the fault inception asserts again at the inception of the second fault, which delays the fault clearance as it requires to wait again for the blocking relay pickup and further logic. At last, as an operational concern, after years of application, the scheme often switches lockout itself due to the poor contact conditions of voltage selection relays or the power module or CT/VT module failure of busbar protection relays. However, in terms of busbar extension or the conversion of feeder panels to interconnector panels, the actual modification of the scheme is with limited complexity.

## 3. High Impedance Busbar Protection (HZBBP) and its Auxiliary Components

HZBBP is a better busbar protection option for new installation in terms of selectivity, sensitivity and speed which requires auxiliary contacts or racking switches, dedicated Class X CTs and trip relays. The fundamental operating principle relies on paralleling the secondary circuits of all CTs to obtain the instantaneous differential $i_\Sigma(t)$ and feed it into a resistor-relay circuit as illustrated in **Figure 3(a).** As CT saturation is a possible and unavoidable event in both external fault and internal fault condition, it either creates a spurious differential which leads to maloperation and trip the whole bus or introduces no differential under internal fault leaving the fault uncleared if overcurrent differential design is used. Therefore, "*high impedance scheme*" is designed with a stabilizing resistor in series with the OC relay to divert spurious differential to the saturation branch without operating. The stabilizing resistor is sized to create a shunt voltage larger than the possible shunt under maximum through fault condition, as described as the lower bound in the first equation of (1), so that through fault stability is guaranteed. To avoid introducing a high voltage to the shunt path with large internal fault current possibly leading to thermal or dielectric failure, the scheme is often designed with a setting resistor in shunt with the high resistance path. The setting resistor reduces the shunt voltage with a smaller equivalent resistance, and it also increases the primary operating current $I_{POC}$ as shown in the second equation in (1). Also, fast operating OC relays is employed to effectively clear the internal fault at the initial fault stage (stage 2 in of internal fault in [10]) before the MOV induction and time-to-saturate of the CTs under severe internal fault. The shunt voltage for setting $V_{SH}$ is also designed to be smaller than 1/3 of the smallest knee point voltage for the parallel CTs as described in the upper bound of the first equation in (1). A CT supervision relay, which automatically shorts the CTs to avoid maloperation under open CT condition, is also included as shown in **Figure 2(a)**.

$$\begin{cases} I_{TH}(mR_L + R_{CT}) \le V_{SH} \le \frac{1}{3}V_K \\ I_{\max} \le I_{POC} = nI_e + I_R + \frac{V_{SH}}{I_{SH}} + I_{MET}(V_{SH}) \le kI_{F,\min} \end{cases} \quad (1)$$

In sum, shunt voltage setting threshold $V_{SH}$ is a critical design consideration for HZBBP [9-10]. The pickup setting is *lower bounded* by the maximum voltage developed across the shunt path under through fault condition with full CT saturation and is *upper bounded* by a certain portion of knee point voltage to ensure operative before the CT get fully saturated under internal fault conditions. The primary operating current $I_{POC}$ is large enough such that the scheme should remain stable and switch out itself with the CTS path under CT open condition of any feeders.

A metal-oxide varistor (MOV) across the high impedance circuit is introduced to clamp excessive voltage to protect both the relay and the circuitry under severe internal fault. The stabilizing resistor and setting resistor should also be able to dissipate energy from the maximum RMS current under a severe internal fault for the expected maximum fault clearance time (e.g. 0.5s) and the resistors should also survive under continuous flow of maximum circuit current in secondary for longer than the switch out timer (e.g. 10s).

With a double bus arrangement, racking switches or auxiliary switches for busbar isolators is introduced, such that the current can enter a correct zone for differential calculations and indicate a correct position for the circuit for tripping. Racking switches are installed inside the bus shutter to indicate the CB position. Accidents occurred as indicated in IEEE C37.234 [6] that the racking switch failed to close completely or was in high contact resistance such that the CTs was open circuit, leading to damage of the terminal blocks and the racking switches in practices.

Under ideal condition, the busbar auxiliary should allow secondary current flow as the same timing as the primary. It is, however, not the case as arcing takes place before the actual mechanical positioning. Besides, if any auxiliaries do not reflect in accurate position, the differential could lead to busbar protection lockout with the CTS element. One should check the integrity of the busbar auxiliaries by performing contact resistance test with the minimum disconnection of wirings. If the double bus arrangement allows on-load transfer, i.e. to switch the circuit from one bus to another bus without de-energizing the circuit, the auxiliary switches should be in early-make-late-break type such that the secondary current flows as the primary without causing any spurious differential. Under such condition, the two buses are coupled through the auxiliaries, and any fault occurs during such operation can lead to tripping of both buses. It was noted that a busbar isolator fault intrinsically is a bus coupler fault as the measuring zone itself does not provide any bus replica information to define the zone and both buses are cleared for further site investigation to determine fault location before a bus to be restored.

Digital relays can be used as the OC element in the high impedance circuit. However, the performance varies with the filtering algorithm inside. To avoid transient noise and short duration unipolar pulse caused with in-zone arrester, three kinds of magnitude can be used [10] –

- **Phasor based Fourier Filter** (Band Pass)**:** Immunity to harmonics and DC offsets, Effective for steady-state fault detection.
- **Raw samples with MAX and MIN:** Avoid under-estimation caused by band pass with low class CTs, Improved sensitivity with better estimation on actual signal magnitude.
- **Waveform recognition with Saturation:** Provide fast fault detection (sub-cycle), Improved robustness against CT saturations by relying on voltage, Incorporated arrester logic to differentiate arrester pulse from bipolar fault signal.

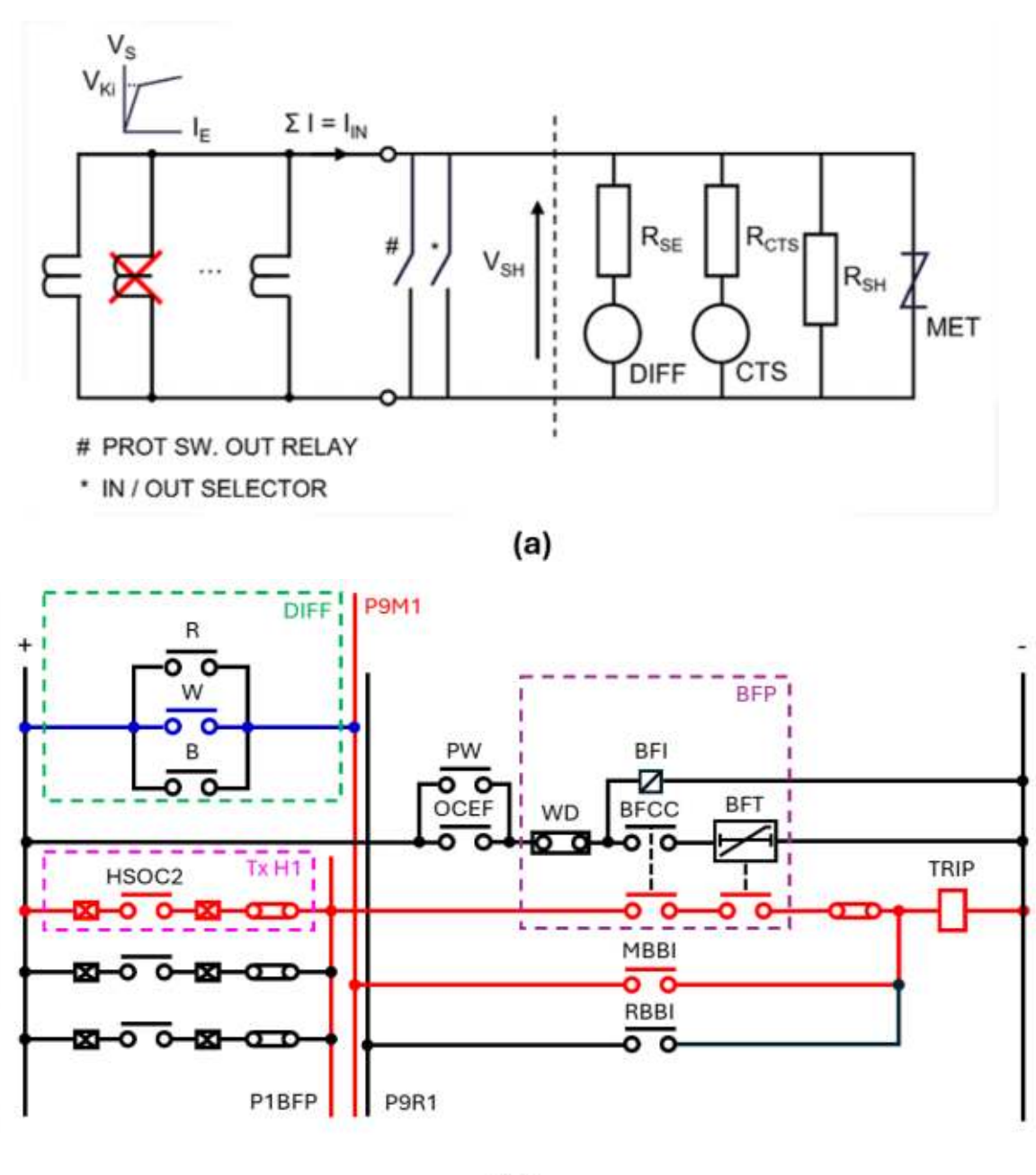

**Fig. 3.** A Possible Implementation of -
(a) AC Schematics of HZBBP,
(b) DC Schematics of HZBBP with BFP

To obtain the summed current and allow neutral current flow under transient and unbalance condition, secondary ring buses are looped through the LV compartment of the switchgear according to the bus position, and the neutral bus is to loop through all panels. The differential is then obtained from a node on the ring bus and routed to the busbar protection panel to connect to the high-impedance circuit. In the absence of a clearly defined ferrule system to indicate the actual wiring, significant challenges arise when breaking the ring under load for busbar extension or during routine maintenance, particularly in locating points of CT low insulation. Additional complications occur in incorrect wiring identification; for example, crossing two different CTs or missing a neutral link can result in maloperation of the protection scheme.

Unlike ILOC, which intrinsically provides a BFP function when the blocking path is restored and the BFP only trips the source, an additional path to provide BFP function as a local backup, possibly as the one shown in **Figure 3(b)**, is included. The incorporation of BFP function is determined by the specific operational requirement of the utilities.

BFP is initiated with the circuit protections as a breaker failure initiation (BFI). When the BFI is activated, and the current check (BFCC) is asserted over the timer BFT, the BFP function is activated to energize the trip bus of the corresponding busbars as a back trip function. To improve the reliability of the scheme, one can include a watchdog (WD) contact to avoid the scheme operation under relay failure condition, and an overcurrent guard (HSOC2) at the HV side of the source transformers. Some schemes as stipulated in IEEE C37.119 would include a slight delay between re-trip and back-trip to provide better selectivity [11]. As stipulated in Section 7.3 of IEEE C37.119 [11], the loss of independence of BFP and circuit protection with multi-function relays may be avoided by programming BFP (or its timer) function in both primary and secondary relays to eliminate any common-mode failure. The additional requirement for DERs with HZBBP is to block reclosing and enable transfer trip only, as illustrated in [11].

HZBBP is not preferrable when bus selection is a compulsory requirement as open CT is possible. EBBP or even NBBP is a possible solution in which the current is sampled with an interposing CT or even numerically.

## 4. From Electronic Busbar Protection (EBBP) to Numerical Counterpart

EBBP, which was an alternative product in history, employed static relays and electronic circuits to perform fault detection. The fault detection algorithms varied in different manufacturers. This included bias differential with diode circuits, comparators and integrators from Siemens, directional aided with differential with saturators and logic gates to identify internal fault from BBC, and ratio differential by phase comparison which creates "restraint blocking" from Mitsubishi. The logics inherit in the modern bus differential relays to be discussed in **Section 5**. In this section, the bias differential scheme coined as RADSS with high through fault stability due to the nonlinearity of diode circuits and the behavior of fault currents under internal fault with unloaded lines (X >> R) and external fault with CT saturation (R >> X) introduced by ABB, which is still in service, is discussed.

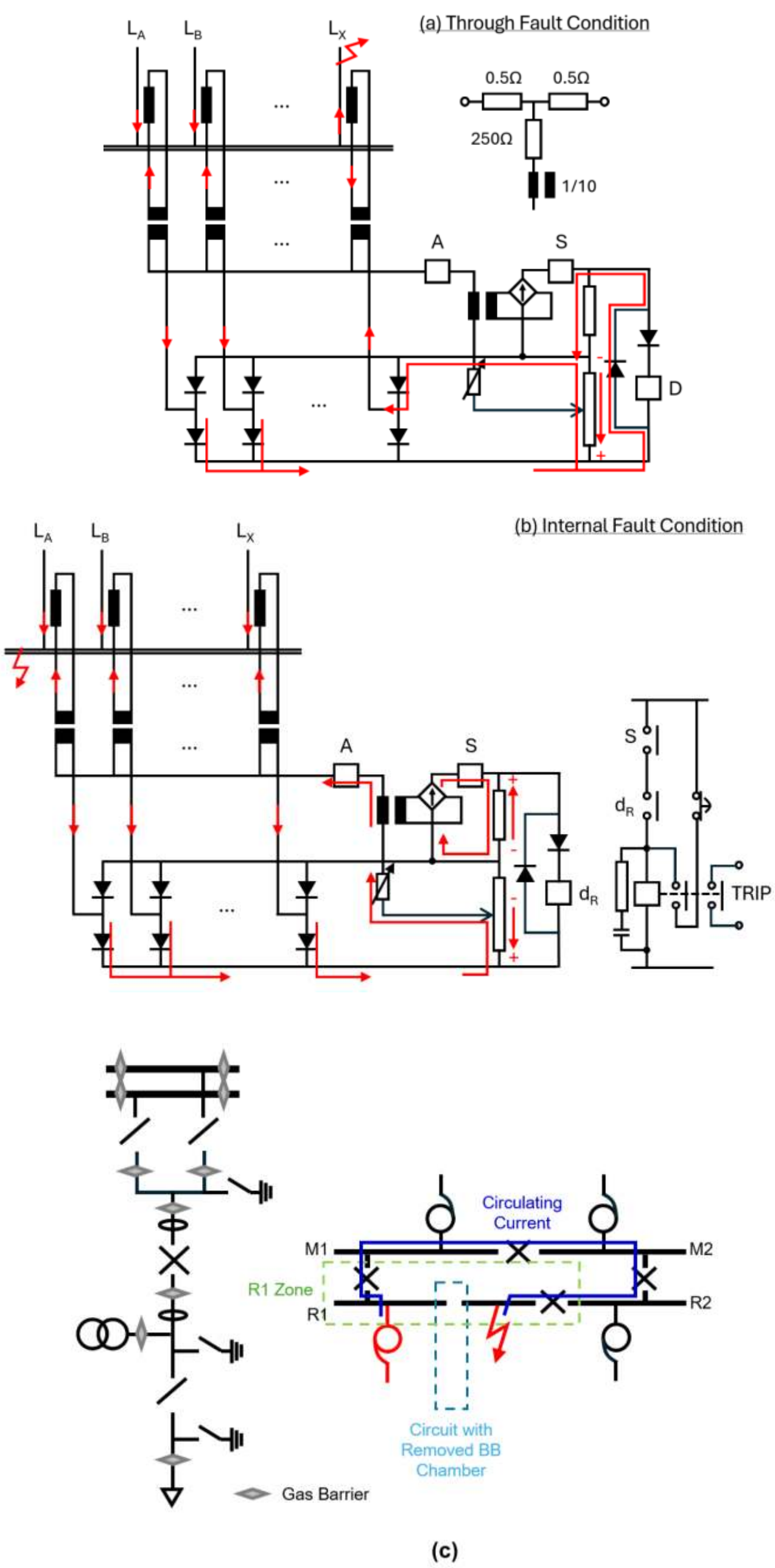

**Fig. 4.** (a) Through Current Path in RADSS; (b) Differential Current Path in RADSS; (c) Circulating Current Issue in IEEE C37.234

**Table 1:** Comparison between Different Busbar Protections in Operating Principles and Performance

| Scheme | | Operating Principle and Trip Logic | Open CT | CT Sat | Expand -ability | Renewable Integration | CT Switching | Evolving Fault | Remarks (e.g. Possible Failure) |
|---|---|---|---|---|---|---|---|---|---|
| LTF | | OC between frame earth and true earth | X | X | High | ✓ | X | X | Fail to detect three phase fault, fail under insulation degradation, possible instability under CEB fault with flashover |
| ILOC | | DOC interlocked with downstream OC blocking | X | X | Medium [1] | Problem arises | X | Delayed Op | Problems on overload and stalling current; Depends on "load" and "source" |
| HZBBP | | Differential OC (DTL) | ✓ | Limited shunt volt | Low [2] | ✓ | Possible Open CT | Possible Lockout | Requirement on CT matching, Class X CT preferred |
| EBBP | RADSS [12] | Biased differential + OC Starter;<br>With restraint blocking under through fault | ✓ | ✓ [3] | Low [4] | ✓ | Switched by Relay | Restraint Delay [5] | Possible electronic device failure, Inaccuracy at setting |
| NBBP | ABB [13] | $i_D[n] = \lvert\Sigma_j i_j[n]\rvert$ "differential"<br>$i_R[n] = \max\{\Sigma_j i_j^+[n], \lvert\Sigma_j i_j^+[n]\rvert\}$ "through"<br>DZ & CZ & RMS OC Check<br>CT Supervision | ✓ | ✓ | | | | EFD Unblock | To mimic the current definition of RADSS; EFD = "$I_{IN} = I_{OUT}$ at the beginning of external fault"; CT Supervision can block or supervise (= high set differential operate level) |
| | TOS [14] | $i_D[n] = \lvert\Sigma_j i_j[n]\rvert$ ; $i_R[n] = \Sigma_j \lvert i_j[n]\rvert$<br>DZ & CZ & ΣIZ<br>Block by WDE as EFD ($\Delta I_D < k\Delta I_R$) | by high set setting | by DPFC detector (blocking) | | | | EFD Unblock | ΣIZ as restraint overcurrent guard; DIFSV as differential supervision |
| | NR [15] | $i_D[n] = \lvert\Sigma_j i_j[n]\rvert$ ; $i_R[n] = \Sigma_j \lvert i_j[n]\rvert$<br>$\Delta I_D[n] = \lvert\Sigma_j \Delta I_j[n]\rvert$ ; $\Delta I_R[n] = \Sigma_j \lvert\Delta I_j[n]\rvert$<br>DZ & CZ & Starter (DPFC of $I_R$ or $I_D$ OC)<br>Block by CT Saturation and CT Superv. | ✓<br>$I_D > [I_{CTS}]$ | by Adaptive Setting [6] | | | | Restraint Delay | Include state estimation to check on open CT; CT Supervision can lock out BBP (require manual switch IN) |
| | SIE [16] | $I_D[n] = \lvert\Sigma_j I_j[n]\rvert$ ; $I_R[n] = \Sigma_j \lvert I_j[n]\rvert$<br>$I_{R,MOD}(s) = \frac{1}{1+s\tau} I_R(s);\ \ I_D > kI_{R,MOD}$<br>NOT EFD & DZ (instant) or<br>NOT EFD & DZ (filtered) + CZ | ✓<br>$I_D > [I_{CTS}]$ | by Zero-Crossing & restraint filtering | Medium (input dependent) | ✓ | X | Restraint Delay | Filtered quantity (BPF) blocked by EFD with $dI_{RES}/dt$; cross stabilization with max ($I_{PH}$), under external fault, instantaneous value algorithm blocks filtered quantity for 150ms; |
| | GE [17] | $I_D[n] = \lvert\Sigma_j I_j[n]\rvert$ ; $I_R[n] = \max \lvert I_j[n]\rvert$<br>Differential OC (unbiased, BPF) or<br>Differential (biased) & Directional | by high set setting | by EFD & Differential Setting | | | | EFD Unblock | No Check Zone, CT Saturation Detector = $\Delta I_D/\Delta I_R$ > k; Directional Element: (External Fault) $\mathbf{Re}\langle \dot{I}_k[n], \sum_{j\neq k} \dot{I}_j[n]\rangle < -\varepsilon$; Allow sensitize for high resistive ground systems |
| | SEL [2] | $I_D[n] = \lvert\Sigma_j I_j[n]\rvert$ ; $I_R[n] = \Sigma_j \lvert I_j[n]\rvert$<br>Differential (biased) & Directional & NOT EFD | by high set setting | By saturation detector [7] | | | | EFD delayed unblock | Directional Element: (External Fault) $\mathbf{Re}\langle \dot{I}_k[n], \sum_{j\neq k} \dot{I}_j[n]\rangle < -\varepsilon$<br>EFD will high set differential element and add a delay of 2 cycles before issuing a trip |

[1] ILOC: Require breaking and reconnect blocking path for feeder panel and perform additional testing on VTS and "forward" blocking at source panel.
[2] HZBBP: Require opening the ring loop of CTs and connecting the trip bus, P1BFP; Excitation current at all buses $I_{POC}$ calculation, which should be bounded by a certain portion of minimum fault current, or to be reduced by increasing $V_{SE}$ if allowed.
[3] EBBP: For internal fault with CT saturation, EBBP requires to operate before saturation within 2 – 3ms; For external fault with CT saturation, the stability characteristics S requires a differential of 0.5 – 0.8 times of total current. The differential current must introduce a voltage larger than that of current through bypass diode to trigger a larger differential flow. It is known as restraint blocking.
[4] EBBP Expandability: Require available diode bridge inlet (~ 13 ways per bus), additional trip and BFP relay which is already obsolete.
[5] EBBP Evolving Fault: The through fault current introduces a restraint blocking voltage in which requires a larger differential to flow through the differential path to create a large enough differential voltage to create a forward bias voltage to the blocking diode.
[6] NBBP (NR) CT Saturation: Low Set (K = 0.2) by Adaptive Weight Anti-Saturation (WA) with DPFC and High Set by Harmonic Blocking
[7] NBBP (SEL) CT Saturation: Recognize the pattern of CT saturation under External Fault with $(I_D + \Delta I_{SAT}, I_R - \Delta I_{SAT})$

Convention: $i[n]$ denotes instantaneous time domain value; $I[n]$ denotes phasor domain value.

RADSS provides a restraint blocking voltage with the "through current" in **Figure 4(a)** to maintain through fault stability and ensure rapid tripping with the "differential current" in the differential path in **Figure 4(b)**, by creating a large differential voltage and a forward bias to the blocking diode to activate the trip element. The stability characteristic is defined by the slope setting $S$, such that the differential current must exceed $S$ times (typically 0.5 – 0.8) the *total incoming current* to initiate operation. For internal bus faults, the protection must operate within 2 – 3ms before CT saturation occurs. The unloaded CTs contribute a large magnetizing inductance as load to reduce the differential voltage. The differential current then exceeds the minimum operating threshold, which ranges from 0.2A to 0.65A, to initiate tripping. RADSS employs impulse-storing circuitry and a self-sealing relay to enable fault detection within approximately 3 ms, even under adverse CT performance. This provides both through-fault stability and internal fault sensitivity, independent of fault level and system time constant [12].

Despite the strong performance of RADSS and other EBBP designs, they are still fundamentally electronic and rely on static relays. They still suffer from inaccuracy in setting, which is to be tuned by compressing the coil in the reed-relay, possible wiring disconnection to be found under maintenance, and component failure under DC surge possibly from battery maintenance. For example, a diode puncture can lead to indiscrimination of differential current and through current or incorrect activation of differential relay.

A busbar isolator fault, as illustrated in **Figure 4(c)**, requires the whole gas zone including the busbar to be removed and separated into two. In such configuration, a fault in the sectionalized bus can lead to a circulating current going in and out the same zone as stipulated in IEEE C37.234 [6] which in turn increases the bias current leading to reduced sensitivity. However, it does not indicate a necessity of low setting the bias, as it is still possible to have fault in the zone without any circulating current.

The diverse protection philosophies employed by the EBBP schemes have been integrated into the multi-functional numerical relays. This integration facilitates rapid clearance of internal faults and maintains stability under different normal operating conditions.

## 5. Evaluation of NBBP Design against Other Busbar Protection Designs

With the advancement of technologies in modern digital relay, the multi-functional digital relay with NBBP is often an alternative option for busbar protection, especially if different operating principles for two main protections are preferred. However, the design philosophies of the scheme vary in different manufacturers. The NBBP designs, their solutions to the problem discussed such as instabilities to open CTs and CT saturations, expandability, performance with DERs and evolving fault are summarized in **Table 1**. The table was generated based on the information provided on the available relay manual online, which may not be indicative.

Operating in low impedance principles, NBBP often maintains its stability with additional components such as check zone (CZ) differential, in which the instantaneous or phasor sum of all circuits in a check zone, possibly for circuits in both main bus and reserve bus in double bus configuration or all circuits in a substation, is compared with the restraint current. The discriminative zone (DZ) is driven by dynamic bus replica in which a positive DC is used to energize a BI through an N.C. and N.O. contact to determine the zone the current belongs to and the bus that the circuit should issue a trip signal.

NBBP can employ different quantities in DZ and CZ or even in different algorithms parallel running. Some NBBP designs prefer instantaneous value. A rectified sum instead of phasor sum is then used as the restraint current. Siemens, however, prefers using an exponential filtered values in which the design can override instability such as CT saturation and open CT within the "zero moment" [16]. As a relay relying on restraint logic instead of blocking, Siemens also enhances its stability with cross-phase restraint by adding the maximum current of the other phases into the restraint current [16]. Some prefer phasor quantity after discrete Fourier transform (DFT), which is essentially a band pass [2], [16], [17]. GE prefers using the max function to avoid de-stabilizing under through fault CT saturation conditions [17]. Deviation of Power Frequency Component (DPFC, i.e. $\Delta i[n] = i[n] - i[n-p]$, where p is the multiples of samples in one cycle) is also a proper option to detect the actual fault quantity independent of load [15]. ABB insists on using the same differential and through current definition to mimic the behavior of RADSS. Through current defined as the maximum of the sum of positive current and that of negative current is used as the restraint, as compared to the diode driven through current creating the stabilizing voltage in RADSS [12]-[13]. In general, differential or restraint OC guard is provided to maintain stability under open CT conditions. If not, then the differential setting should be high set.

Other than the CZ, external fault detector (EFD) is often an option to maintain stability under through-fault CT saturation conditions. It is often based on the delta quantity. For example, Siemens uses $dI_R/dt$ [16], while Toshiba and GE use $\Delta I_D/\Delta I_R$ as reference [14], [17]. Another possible EFD is the directional element used in GE and SEL [2], [17]. If all current flows into the zone, all current should have a positive phasor product; If there is an external fault, there should exist a reference current in which could lead to a negative phasor product with the others. To reduce computational effort, one can find a reference, possibly the one with largest current to determine the phasor product with the others. The EFD can also act as a blocking element as CT saturation detector. NARI, unlike the others favored using EFD, prefers adaptive setting with adaptive-weighed anti-saturation to detect no saturation, and a harmonic blocking logic to high set the differential [15]. EFD in SEL also provides such adaptive settings to avoid permanent blocking the operation under internal fault [2].

The performance of NBBP to evolving fault often depends on the existence of EFD and the corresponding delay and additional restraint (a.k.a. restraint philosophy) to maintain stability under evolving fault. It should be noted that blocking element, in case the restraint current is not additional provided and the operating characteristic is not high set, should operate faster than the restraint logic upon unblocking [18]. For example, if the EFD in SEL operates, a delay of 2 cycles is provided to ensure the operating point in differential zone before issuing a trip [2]. It helps maintain stability under evolving fault conditions, when external fault evolves into internal fault and CT saturation is not heavy enough to blind the condition. Additional delays for restraint philosophy are caused by the time to move the operating point to restraint area to operating area given that the restraint current is added to with cross-phase stabilization and exponential filter [16].

## 6. Conclusion

The introduction of modern multi-functional digital relays has advanced the implementation of NBBP. However, their performance remains questionable due to variations in operating principles, particularly the blocking logic and restraint algorithms adopted by different manufacturers.

To improve the performance of busbar protection under the possible changes in busbar configuration, CT locations, degraded performance of auxiliaries and transducers, stochastic performance of mechanism and dielectrics, the following improvements can be considered [10].

- Distributed busbar protection architecture with IEC61850 - to replace copper wiring with digital messaging to improve reliability and enhance interoperability, but to avoid any failures with ageing components leading to prolonged fault restoration
- Enhanced differential protection algorithm – Adaptive restraint logics, saturation / external fault detectors, dynamic thresholding often maintains the stability under through fault conditions. Yet the overlapping in blocking and restraining design may lead to prolonged or failure in fault clearance. Further studies is required to ensure the performance under critical fault conditions, such as evolving fault, high resistance fault, in-zone component failure and fault at different inception angles.
- Improved CT supervision and redundancy – Multi-criteria CT failure detection, redundant CT inputs, cross checking between bays (such as state estimation from NR's relay [15]) can be used to monitor and early discover any CT failure conditions to maintain busbar protection stability. It also checks the integrity of transducers of the NBBP relays.
- Automatic Dynamic Busbar Zone Configuration – Reconfiguration based on CB status and busbar topology reduces the risk of incorrect zone assignment can be provided possibly with state estimation. It is essential for complex GIS design (such as those with bypass switches) and multi-section busbar configuration.
- Enhanced BFP and Interlocking Logic – The tight integration between busbar protection, BFP, interlocks and lockout conditions should be well design or coordinated to avoid the same incidents in Taiwan 303 Blackout. It can also reduce fault clearance time and improve selectivity.

This paper has outlined the operating principles and limitations of the mature technologies in industrial applications to supplement the IEEE standard. This paper also provides, for the first time, a systematic comparison of contemporary NBBP schemes against traditional approaches. The results emphasize that even subtle differences in scheme design are critical for ensuring stability under normal operating conditions.

**Mr. Karl M. H. LAI**

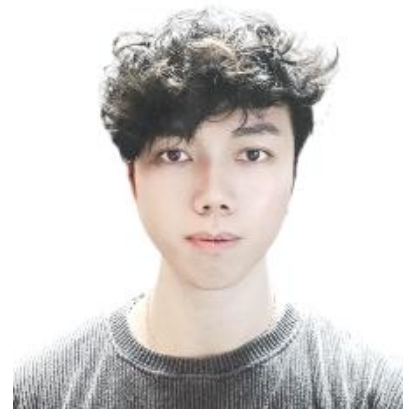

Karl M.H. Lai received his MSc in HKU in 2022. In 2019, he joined CLP as a protection engineer specializing in system, generator and IBR protection and control. He has a particular interest in advanced control, filtering and estimation in IBR and HVDC systems.

**Prof. Yunhe HOU**

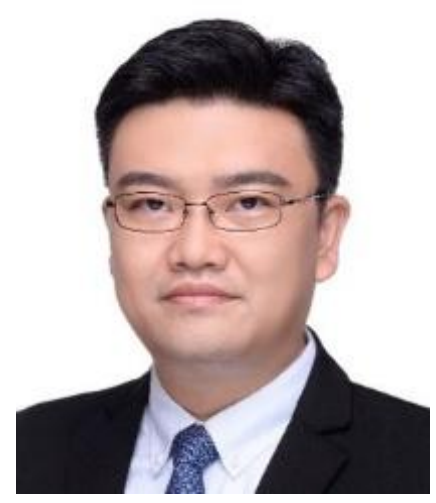

Yunhe Hou is an Associate Professor with Department of Electrical and Computer Engineering, HKU. He received his PhD in 2005 from Huazhong University of Science and Technology, China. He specializes in power system resilience and planning and optimization in power systems.

**Mr. Kwunhang WONG**

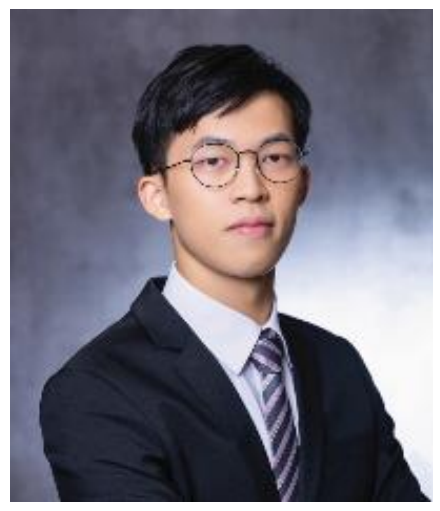

Kwunhang Wong is a PhD candidcate in Department of Electrical and Computer Engineering, HKU. He specializes in neuromorphic computing, hardware-software co-design and differential privacy.